\documentclass[a4paper,20pt]{article}
    \usepackage[T1]{fontenc}
    \usepackage{graphicx}
    \usepackage{epsfig}
    \usepackage{amsmath}
    \usepackage{amsfonts}
    \renewcommand{\abstract}{}
\begin{document}
\makeatletter
\renewcommand{\@oddhead}{\textit{YSC'14 Proceedings of Contributed Papers} \hfil \textit{B.Wszo\l ek}}
\renewcommand{\@evenfoot}{\hfil \thepage \hfil}
\renewcommand{\@oddfoot}{\hfil \thepage \hfil}
\fontsize{11}{11} \selectfont

\title{Puzzling Phenomenon of Diffuse Interstellar Bands}
\author{\textsl{B. Wszo\l ek}}
\date{}
\maketitle
\begin{center} {Jan D\l ugosz Academy, Institute of Physics, al. Armii
Krajowej 13/15, 42-200 Cz\k{e}stochowa , Poland \\ bogdan@ajd.czest.pl}
\end{center}

\begin{abstract}
The discovery of the first diffuse interstellar bands (DIBs) dates
back to the pioneering years of stellar spectroscopy. Today, we know
about 300 absorption structures of this kind. There exists a great
variety of the profiles and intensities of DIBs, so they can not be
readily described, classified or characterized. To the present day
no reliable identification of the DIBs' carriers has been found.

Many carriers of DIBs have been proposed over the years. They ranged
from dust grains to free molecules of different kinds, and to more
exotic specimens, like hydrogen negative ion. Unfortunately, none of
 them is responsible for observed DIBs. Furthermore, it was shown that
a single carrier cannot be responsible for all known DIBs. It is
hard to estimate how many carriers can participate in producing
these bands. The problem is further complicated by the fact that to
this day it is still impossible to find any laboratory spectrum of
any substance which would match the astrophysical spectra.

Here, a historical outline concerning DIBs is followed by a brief
description of their whole population. Then, a special attention is
focused on the procedures trying to extract spectroscopic families
within the set of all known DIBs.
\end{abstract}

\section*{Introduction}
\indent \indent The presence of diffuse absorption features in the
optical spectra of reddened stars has been known for many years (see
\cite{herbig1995} for extensive review). DIBs were firstly mentioned
by Heger \cite{heger} and then confirmed by Merrill
\cite{merrill1934}. Their interstellar origin was established on the
basis of correlations between their strength and parameters of the
dust or gas, such as reddening or atomic hydrogen column density.
The name `diffuse interstellar bands' is given to all discrete
features, observed in the spectra of reddened OBA stars, which
remain unidentified.

Some 300 DIBs are known in total, spanning the wavelength range from
0.4 to 1.3 $\mu$m. The most widely observed and discussed DIBs
include those at 4428, 5780, 5797, 6177, 6196, 6203 and 6284 \AA. By
convention, each DIB is identified by its central wavelength in \AA
\,to four significant figures. One exception is that at 4428 \AA,
amongst the first to be studied, which traditionally has been
rounded to 4430 \AA. DIBs differ greatly in shapes from very broad,
e.g. 4430, to very narrow, like 6196 \AA. Many of the bands are
extremely weak; their central depths do not exceed 1\% of the
continuum in the majority of the observed stars. Precise
measurements of such bands are still very difficult and the
published research papers deal usually with the strongest DIBs. The
adjective `diffuse' may be slightly misleading, especially in the
case of relatively sharp structures. DIBs are usually sharper than
stellar lines observed in the same spectra. Nevertheless, the name
is justified by the fact that even the sharpest DIBs are still
broader than interstellar atomic, ionic or molecular features. The
broadening of the profiles of DIBs is presumably due to unresolved
rotational structure, possibly compounded by lifetime broadening of
the upper states.

During the past years the interest for DIBs has grown considerably,
particularly because the new observing techniques and the improved
quality of the spectra allowed a deeper analysis of their profiles,
highlighting more and more details on their behaviour and therefore
making them interesting candidates as markers of the physical and
chemical status of the interstellar medium. However, in spite of the
higher resolution and the excellent high signal to noise ratio which
can be obtained from modern spectrographs coupled to CCD detectors,
the nature of the carriers of the DIBs remains a mystery.

\section*{The discovery of DIBs}
\indent \indent In 1922 Heger \cite{heger} reported the discovery of
two spectral features, centred near 5780 and 5797 \AA, in spectra of
some spectroscopic binaries. These discovered features were
considered as interstellar ones, however, it was only in the 1930s
when works of Merrill (e.g. \cite{merrill1934, merrill1938})
confirmed this hypothesis. Merrill demonstrated that these puzzling
features did not participate in the velocity variations of
spectroscopic binaries and furthermore their strength increased with
distance and with the degree of the interstellar reddening of the
star that furnished the background continuum. That evidence was
reinforced by the work of Beals and Blanchet \cite{beals}, and
especially by the extensive study of the broad feature centred at
4428 \AA\ \cite{duke}.

Bright, near-by stars obscured by one cloud would be the most
appropriate candidates for the study of DIBs, however, number of
such stars is strongly limited. It is essential to mention that
accessible early types stars are usually either very distant or very
slightly reddened, thus the molecular features of spectra are either
formed in several clouds along any sightline or are too weak to be
measured with proper precision. When several clouds are situated
along a sightline, researches get the average spectra and any
interpretation of such spectra is more difficult.

The next problem is contamination of interstellar spectra with
telluric lines. They are lines and bands of the atmospheric origin.
Many of DIBs occur in regions masked by atmospheric $O_2$ and $H_2O$
lines, so overlying atmospheric structure has to be properly
removed. To do so, one has to divide a given spectrum by a spectrum
of the standard. As a standards, the unreddened stars, like e.g.
$\alpha$Peg or $\alpha$Cyg, are used. Cancellation of atmospheric
features allows new DIBs to be discovered.

The number of known DIBs keeps growing. The first survey of DIBs was
published in 1937 \cite{merrill1937}. In the year 1975 the major
survey of DIBs was published by Herbig \cite{herbig1975}. It
contained 39 DIBs (twenty of them were observed for the first time)
detected in the range of 4400-6700 \AA. All these features were
clearly seen in the spectrum of the heavily reddened star HD183143.
The replacement of photographic emulsion by solid state detectors
for stellar spectroscopy has resulted in the detection of many new
DIBs. In his new survey Herbig \cite{herbig1995} placed over 130
features and Kre\l owski et al. \cite{krelovsky1995} added to this
wealth of data yet another 52 weak DIBs. Galazutdinov et al.
\cite{galazutdinov2000a} presented an atlas of 271 DIBs between 4460
and 8800 \AA, of which more than 100 are new bands. Set of few
features known in 1930s got much bigger and now the number of DIBs
is around 300. However, existence of some of them is not
sufficiently proved. On the other hand, many features probably still
wait for their discovery to come.

\section*{The problem of DIBs' carriers}
\indent \indent The identity of the DIB carrier(s) is a long-standing problem that has
simultaneously fascinated and frustrated researchers for very long time.
 The various proposals are reviewed in detail by Herbig \cite{herbig1995}. Although
 numerous, the DIBs are weak and the sum of their absorptions is very
small, e.g. in comparison to the 2175 \AA \,feature on the
extinction curve (e.g. \cite{whittet}). Thus, the absorbers need not
be very abundant. The big number of known DIBs, and their widespread
distribution across the optical spectrum, strongly suggest that more
than one carrier is involved. Single species of forbidding
complexity would be needed to account for all of them
\cite{herbig1975}. Further support for multiple carriers arises from
intercorrelations of the features with each other and with
reddening, suggesting the existence of several `families' (e.g.
\cite{krelovsky1987}, \cite{moutuu}).  Origins in both dust grains
and gaseous molecules have been proposed. Features produced by
solid-state transitions in the large-grain population should exhibit
changes in both profile shape and central wavelength with grain
size, and emission wings would be expected for radii \textgreater
0.1 $\mu$m \cite{savage};  no such effects have been observed. There
is also a lack of polarization in the features that might link them
to the larger aligned grains. If the carriers are solid particles,
they must be very small compared with the wavelength. The
possibility of a small-grain carrier for the DIBs may be examined
further by searching for correlations between their strengths and
parameters of the UV extinction curve. The ratio of equivalent width
to reddening provides a convenient measure of DIB production
efficiency per unit dust column in a given line of sight. Typically,
this ratio displays a weak positive correlation with the
corresponding relative strength of the 2175 \AA \,bump and a weak
negative correlation with the amplitude of the FUV extinction rise
\cite{witt}. These results clearly fail to establish any firm
associations: on the contrary, it can be concluded that DIB carriers
and the bump carriers are not directly related, as the bump is less
susceptible to variation  than the DIBs and still present in lines
of sight where the DIBs are negligible \cite{benvenuti}. The
observations merely suggest that there is some correlated behaviour
in their response to
 environment.

There has been a degree of consensus in the recent literature that
the most plausible candidates for the DIBs are carbonaceous
particles that might be classed as very small grains or large
molecules - specifically, PAHs and fullerenes
\cite{galazutdinov2000b, herbig1995}. Ionized species are favoured
over neutral species as they have stronger features in the visible.
Observations show that the DIBs become relatively weak inside dark
clouds \cite{adamson}, consistent with a reduction in the abundance
of the carriers in regions shielded from ionizing radiation. The
weak negative correlation found between DIB strength and the
amplitude of the FUV rise might be explained if they are produced by
ionized and neutral PAHs, respectively \cite{desert}. However,
identification of specific DIBs with specific species is
problematic, as the techniques used to study their spectra in the
laboratory introduce wavelength shifts and line broadening
\cite{salama} and this severely hinders comparison with interstellar
spectra. As early as the 1930s, the question was raised whether
individual interstellar clouds have identical spectra and, thus,
identical physical parameters. Herbig and Soderblom
\cite{herbig1982} demonstrated convincingly that the profiles of
DIBs can be modified by Doppler splitting. These observations raised
the challenge of determining intrinsic profiles of the DIBs free of
any Doppler splitting in order to compare them with laboratory
spectra. Such profiles can be observed only in single clouds. Kre\l
owski and Walker \cite{krelovsky1987} showed that the relative
strengths of 5780 and 5797 DIBs were quite different in the spectra
of different stars. This demonstrated that the DIBs do not have a
common origin and individual clouds may differ both in DIB intensity
ratios and in the shapes of the extinction curves produced by
interstellar dust.

\section*{Spectroscopic families of DIBs}
\indent \indent As was already mentioned all known DIBs form very
inhomogeneous sample. Some of them are relatively strong, contrary
to the others which are extremely weak. There are DIBs which are
narrow (e.g. 5797), and there are very broad bands (e.g. 4430). This
morphological heterogeneity of observed DIBs indicates, even without
invoking the other arguments, that there are many carriers of DIBs.
In 1980s researchers started their tries to isolate families of DIBs
upon their morphological features (e.g. \cite{jozafatson}, Snow
1987). It is expected that a progress will be possible, and some
carriers will be closer to be identified, when all known DIBs are
divided into spectroscopic families in such a way that only one
carrier is responsible for all bands belonging to a given family
\cite{wszolek}. The dividing of such kind is to do only by analysis
of astronomical spectra. But the task is not easy at all. The main
obstacle to isolating spectroscopic families of DIBs is the noisy
correlation. Owing to the very high level (as the data analyzed by
Wszo\l ek and God\l owski \cite{wszolek} reveals) of noisy
correlation in observational data, the ability of the statistical
methods to isolate spectroscopic families is very limited. The tight
linear correlation, expected between members of the same
spectroscopic family, is effectively hidden by noisy correlation and
the measurement errors.

\section*{Discussion}
\indent \indent Within the Milky Way, DIBs have been observed
towards more than a hundred stars and there is still no definitive
identification of DIB carriers. Recent studies indicate that the
environmental behaviours of DIBs reflect an interplay between
ionization, recombination, dehydrogenation and destruction of
chemically stable species. It is therefore of interest to study DIBs
in different environments, including external galaxies where many
DIBs were also found (e.g. \cite{sollerman}).

From the other side, interdisciplinary study of the problem seems to
be of crucial importance here. Astrophysically oriented laboratory
experiments are probably the only way to identify carriers of DIBs.
85 years after the first discovery of DIBs one has to agree that the
full solution of intriguing problem of their carriers is still
ahead.

\end{document}